\documentclass[12pt, a4paper]{article}
\usepackage[pdftex]{graphicx}
\usepackage[margin=1in]{geometry}

\usepackage{tikz}
\usepackage{natbib}
\usepackage{anyfontsize}

\usepackage{subcaption}

\usepackage{amsmath}
\usepackage{titlesec}

 \usepackage{float} 

\usepackage{graphicx,wrapfig,lipsum}
\usepackage{wrapfig}
\usepackage{rotating}
\usepackage{pslatex}
\usepackage{colortbl}

\usepackage{setspace}
\usepackage{lscape}
\usepackage{multirow}
\usepackage{ctable}
\usepackage{amsmath}
\usepackage{eurosym}
\usepackage{booktabs}
\usepackage{dcolumn}
\newcolumntype{.}{D{.}{.}{-1}}
\usepackage[toc,page]{appendix}

\usepackage{graphicx}

\usepackage{multicol,caption}

\usepackage{color}
\definecolor{darkred}{rgb}{0.5,0,0}
\definecolor{darkgreen}{rgb}{0,0.5,0}
\definecolor{darkblue}{rgb}{0,0,0.5}

\usepackage{hyperref}

\hypersetup{colorlinks,
 linkcolor=red,
 filecolor=darkgreen,
 urlcolor=darkblue,
 citecolor=darkblue,
}

\graphicspath{ {./figures/}  }

\begin{document}

\title{
Learning to Predict with Highly Granular Temporal Data: Estimating individual behavioral profiles with smart meter data\thanks{
This work is funded by the UK ESRC Consumer Data Research Centre (CDRC) grant reference ES/L011840/1}
 }

\author{
Anastasia Ushakova  
\\
  University College London\\
  \href{mailto:anastasia.ushakova.14@ucl.ac.uk}{anastasia.ushakova.14@ucl.ac.uk}
\and
Slava J. Mikhaylov \\
University of Essex\\
 \href{mailto:s.mikhaylov@essex.ac.uk}{s.mikhaylov@essex.ac.uk} \\}

\maketitle

\begin{abstract}
\noindent 
Big spatio-temporal datasets, available through both open and administrative data sources, offer significant potential for social science research. The magnitude of the data allows for increased resolution and analysis at individual level. While there are recent advances in forecasting techniques for highly granular temporal data, little attention is given to segmenting the time series and finding homogeneous patterns. In this paper, it is proposed to estimate behavioral profiles of individuals' activities over time using Gaussian Process-based models. In particular, the aim is to investigate how individuals or groups may be clustered according to the model parameters. Such a Bayesian non-parametric method is then tested by looking at the predictability of the segments using a combination of models to fit different parts of the temporal profiles. Model validity is then tested on a set of holdout data. The dataset consists of half hourly energy consumption records from smart meters from more than 100,000 households in the UK and  covers the period from 2015 to 2016.  The methodological approach developed in the paper may be easily applied to datasets of similar structure and granularity, for example social media data, and may lead to improved accuracy in the prediction of social dynamics and behavior.

\vspace{0.5cm} \noindent \textbf{Keywords}: big data,  time series models, consumer behavior, smart meters
 
\end{abstract}

\newpage

\doublespacing
\section*{Introduction}

Social and political science research on time series data often focuses on prediction models that strive to understand the dynamics behind the outcome variables \citep{box-steffensmeier_freeman_hitt_pevehouse_2014}. Time series classification models have been used for predictions of conflict and civil war \citep{muchlinski2015comparing, hegre2013predicting}, political instability \citep{goldstone2010global} or design of aggregated indicators, such as political accountability \citep{chappell1985new}. 

Fine-grained, complex non-stationary time series that characterize social dynamics have entered political science research via social media analysis \citep{digrazia2013more}. Prediction of individual behavior with such data is challenging. A mixture of methods to predict behavior with such data is posed as an alternative.  The aim for this paper is to consider challenges and opportunities that may be associated with large highly-granular temporal datasets. In particular, the focus is on challenges relating to aggregation of time-series in the context of big data.

Sampling frequency (or equivalently aggregation level) of data affects the output of analysis. Intuitively, it is expected that the aggregation results in some information loss. The question posed is about the importance of this kind of information for inferences about the process. Aggregation is useful for data reduction, which, in turn, speeds up the computation. In general, the information lost in aggregation depends on properties of the process itself. For example, consider a signal that has very gradual variation over time, then increasing the sampling interval, or equivalently increasing aggregation over very granular samples, may not have much impact on inference about the process. Conversely, if activity varies rapidly, then oit can be expected that aggregation will have a large impact and significantly limit any insights obtained from the analysis. In the event-data application, \cite{shellman2004time} demonstrates the varying effect of aggregation on both segmentation and prediction. We expect the effect to be even more pronounced for highly granular time-series big data. 

As an illustrative example, we analyze smart meter data that records household energy consumption at half-hourly intervals. Such data may be used as a proxy for individual household behavior and activities \citep{ANDERSON201758}. In order to make the approach more generalizable, we propose a two stage procedure: first, segment the population with unsupervised clustering methods to categorize behavior; second, predict cluster allocation using only time series features. 

The paper proceeds as follows. First, we introduce our case study based on a sample of smart meter data available for the UK throughout 2014 and 2015. We assesses several methods for clustering time-series with the aim of segregating consumer behavior. This is followed by prediction of behavioral classes from the individual time-series records. The paper concludes with a discussion on how aggregation affects the analysis in our case study.

\section*{Smart Meter Time-Series Data}

Time series data constitutes an ordered sequence indexed by time alongside values of the variables of interest at each point in time. For smart meter data there are various ways to represent such a time series sequence. For instance, one sequence could represent the total consumption per day, while another could track hourly energy consumption. We can then model different data generation processes depending on our level of aggregation. 

\subsection*{Independence Assumption}

In the case of smart meter data, data may be analyzed in either a univariate or multivariate setting. Traditional analysis is mainly univariate in nature and would impose an independence assumption across energy consumption levels if we are interested in fitting a parametric model.  If we are concerned with the prediction of average (or aggregated) energy demand that is composed of consumption by individual users, then correlation or independence between streams may affect the aggregated processes. In the univariate case,  each customer's time series is taken separately and it is attempted to predict their consumption using only their historical behavior.

\begin{figure}[H]
\centering
  \includegraphics[width=0.5\textwidth]{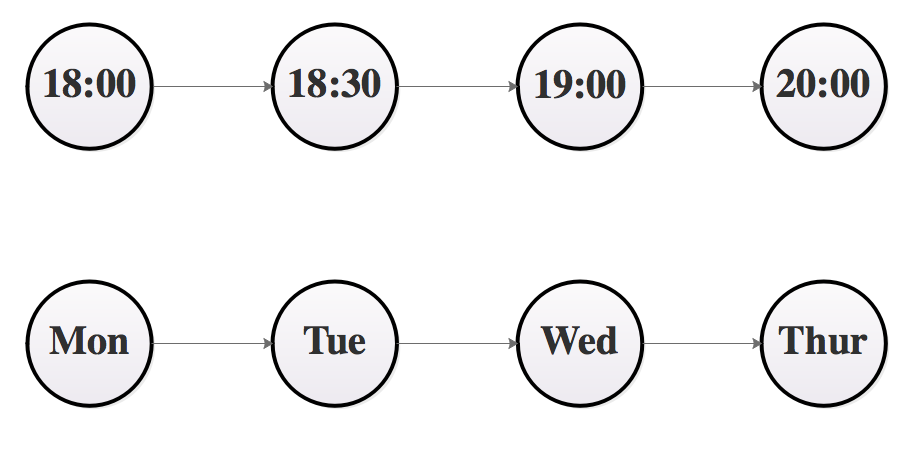}
  \caption{\emph{Chains based on half-hourly readings and total daily readings.}}
  \label{fig:chains}
\end{figure}

Energy consumption may be viewed as a first order chain of preceding readings. Figure \ref{fig:chains} presents a schematic illustration. We assume that there are no interdependencies among the nodes in the chain other than on the previous time-step. Each time period is conditioned on the previous one. However, as an extension a second-order chain may be considered where \textbf{t3} may be dependent on \textbf{t1} and \textbf{t4} dependent on \textbf{t2}. Such models may be generalized to higher-orders at the expense of increasing model complexity and a drop in interpretability.

\subsection*{Description of Data}

A summary of the dataset  is given in Table \ref{tab:all}. From the total set of 8.5 billion observations, we study two sub-samples of smart meter data streams.   Aggregated patterns are calculated by taking the average half-hourly consumption across the whole year for each unique consumer at each postcode sector.  This significantly reduces the volume of data. It is associated with certain levels of variability across the units of analysis, but the variability within the individual customer records is collapsed. To assess how much we can learn about the true dynamics from this aggregated level, we compare the aggregated results to those obtained on a disaggregated raw sample.

The overall dataset is sufficiently large and this may present computational limitations for some analyses. One approach, implemented in the disaggregated sample, is a random draw of 1,100 individuals. For privacy and security reasons, the overall sample is not used in the analysis. We present summary figures below to illustrate underlying data volumes. Computationally, datasets of such sizes can be problematic, especially for methods that extensively use matrix transformations.

\begin{table}[H]
\begin{center}
\centering
\begin{tabular}{ @{} l c c c c c @{} }
\toprule  

Data  & Overall  & Aggregated Sample & Disaggregated Sample \\
\hline \hline
Unique identifiers &  489,000 & 8,171 & 1,100\\
Days & 365 & 365 & 365 \\
Daily readings & 48 & 48 & 48\\
Total observations &  8,567,280,000 & 143,155,920&  19,272,000  \\

\bottomrule
\end{tabular}
\end{center}
\caption{\emph{Data structure.} The structure of our smart meter database and the samples. Note: Aggregated sample is average consumption at each geographical reference level (post-code sector).}
 \label{tab:all}
\end{table}

\begin{figure}[H]
\centering
  \includegraphics[width=0.95\textwidth]{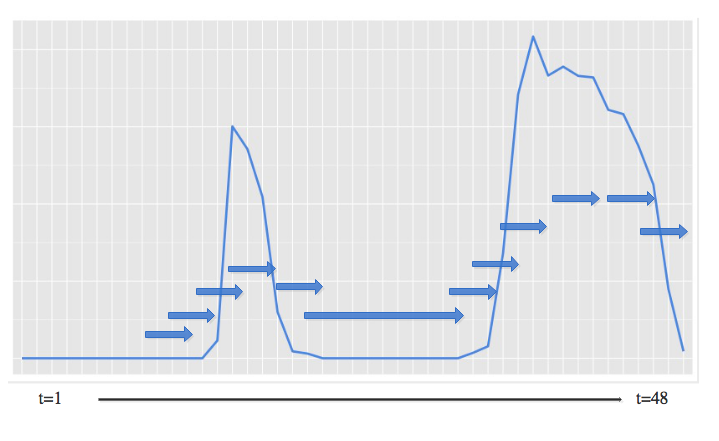}\\
    \includegraphics[width=0.95\textwidth]{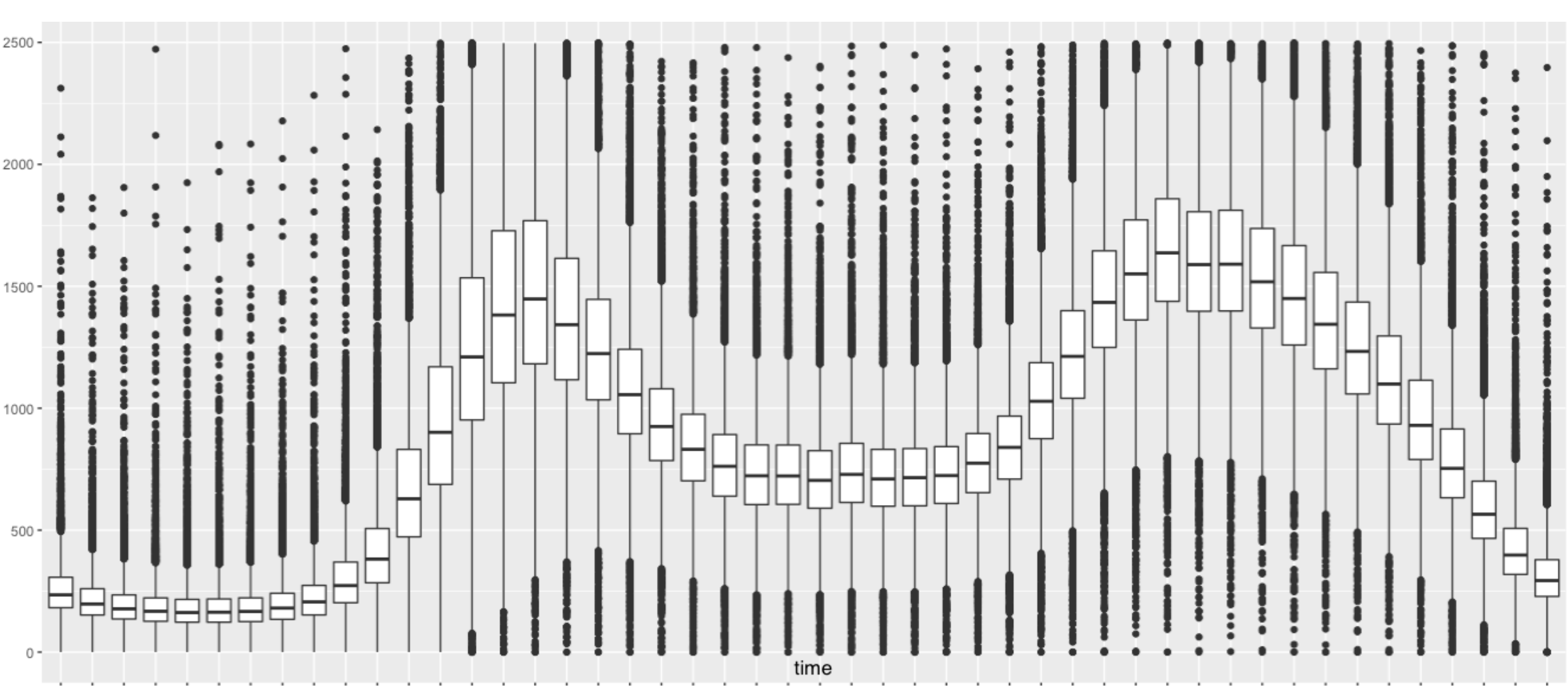}
  \caption{\emph{Decisions on consumption  can be made as granular as at each \textbf{t} to more aggregated structures such as evening, morning peak hour and the time intervals in between.}}
  \label{fig:dec}
\end{figure}

Figure \ref{fig:dec} presents an example of the average daily consumption pattern and variation around this average for a sample of consumers randomly taken from the overall dataset. As may be expected, the shape of consumption behavior aligns with morning and evening peaks. At the same time, if we are to differentiate among the patterns, the variation around the mean and median consumption may generate additional insights about consumer behavior. 

\section*{Methodology}

This section presents the workflow of the analysis.  It is worth noting that a number of other solutions may be considered at the pre-stage of the method (e.g. feature transformations). However, here variable transformations are avoided in order to preserve the interpretability of the analysis as well, as in this case it is important to ensure that the analysis may be replicated using the raw smart meter data without any modifications applied. The significance of this is driven by the applicability of the research method within the industry, for instance.  
  
Our approach is based on a combination of both unsupervised and supervised machine learning techniques. First, the process that may have generated the patterns within the data is studied in order to find a way to group these based on the similarity of that process -- a process often referred to as clustering. Since the data is unlabelled \emph{a priori}, this step is also useful for segmenting large data sets into groups that can then be studied separately. It is often the case that these clusters may be associated with real-world segmentations in the data. 

In the final step we predict assignment to clusters based only on time-series features. This models a setting wherein the researcher may be interested in individual behavior (clusters) without access to any additional information apart from past behavior. We perform this analysis both on the aggregated and disaggregated data streams. 

\section*{Segmentation and Labelling of Time-Series}

As discussed, the dataset represents solely the readings from smart meters and contains no information on individual characteristics of the users and properties. We use unsupervised machine learning techniques to segment the data and create artificial labeling. The next section will assess how well such labels can be predicted from the time-series data. The main goal of this section is to develop a method to read new unseen data and allocate it to a group of already known segments.  Clustering is being accessed as a feasible strategy for segmenting large, granular data. Related work has been undertaken in energy classification using smaller and more aggregated samples \citep{albert2013smart, mcloughlin2015clustering, haben2016analysis}.

\subsection*{Clustering}

Clustering is an unsupervised machine learning method that is used primarily to associate a simplified underlying structure with unlabelled data. For example, in the smart meter case, having solely energy consumption recordings, little is known as to whether the consumption patterns may be aggregated into similarity groups. For instance, people who work full time may be grouped together, while those who are at home throughout the day may also be clustered together. The objective is thus to find an algorithm which ensures that similarity between individuals within each cluster is maximized, while also maximizing dissimilarity between clusters.

To date, a number of methods have been developed for clustering data. While many of these give a reliable performance on static data, they often disregard the dynamic structures of clusters. This poses further challenges if we are to consider the spatial and temporal dimensions in the analysis. One of the immediate solutions could be to transform dynamic data into the static format. For example, we may calculate the mean for each of the individuals and create a numerical indicator that represents an estimate of average consumption for the individuals in our sample. This can also be done for geographical references, reducing the dimensionality of the data and allowing for greater generalization.  According to \cite{liao2005}, the decision on which clustering method is appropriate for time series also depends on the data type. The characteristics can include: discrete vs real valued, uniformity of the sample, univariate vs multivariate series, and lengths of time series considered for the analysis. 

Most clustering algorithms are designed to maximize dissimilarity among the groups using various distance measures (e.g., k-means, hierarchical), while others may consider the underlying data generation process (e.g., Gaussian Mixture Models, Bayesian clustering by dynamics). An important issue for these algorithms is how to treat outliers. For instance, whether outliers are weakly assigned to clusters (with some probability) or are associated strictly with a specific cluster (absolute/hard clustering).

  K-means clustering is the most popular approach due to its simplicity and fast minimization of the similarities among the objects within each class centre. It is well suited for data sets with static features. For highly variable temporal variables, the assignment of the cluster may be highly unstable as individuals are likely to be assigned to a different cluster subject to the day and time. As an alternative, we consider a Gaussian Mixture Model (GMM) based on a probabilistic model \citep{10.2307/2532201}. Such a setting brings about the ability to handle diverse types of data, including dealing with missing or unobservable data that may have contributed to variation differences among segmented groups. This is achieved by assigning a probability to a segmented group membership. Under greater uncertainty about the assignment, additional variables may be introduced or the individual may be treated as an outlier or belonging to an uncertain group. Unlike k-means, it produces stable results and selects the number of clusters using the probability density fit. Clustering results are also replicable and remain the same regardless of how many times we run the algorithm. 


\subsubsection*{Gaussian Mixture Models}

%

Gaussian Mixture Models constitute a probabilistic method for clustering that handles diverse types of data, including dealing with missing data and hierarchical structures.  The probabilities for each data point to be in a particular cluster are first assigned and then  a cluster is allocated to each point using those probabilistic measures. The mixture is formed using the probabilities obtained from the standard Gaussian representation:

\begin{equation}
  P[\boldsymbol{x}|\boldsymbol{\mu},\boldsymbol{\Sigma}]= \frac{1}{\sqrt{2 \pi \mathrm{det}(\boldsymbol{\Sigma})}} \exp \left \{ -\frac{1}{2} (\boldsymbol{x}-\boldsymbol{\mu})^{\top} \boldsymbol{\Sigma}^{-1} (\boldsymbol{x}-\boldsymbol{\mu})\right \}\;,
\end{equation}

 with $\boldsymbol{\mu}$ representing the mean vector and $\boldsymbol{\Sigma}$ being a covariance matrix. A mixture of Gaussians is then represented as the following: 

\begin{equation}
  P[\boldsymbol{x}] = \sum_{i=1}^{H} P[\boldsymbol{x}|\boldsymbol{\mu}_i,\boldsymbol{\Sigma}_i] P[\boldsymbol{x}\in \mathrm{cluster}\; i]\;.
\end{equation}

\begin{figure}[H]
\centering
\includegraphics[width=0.95\textwidth]{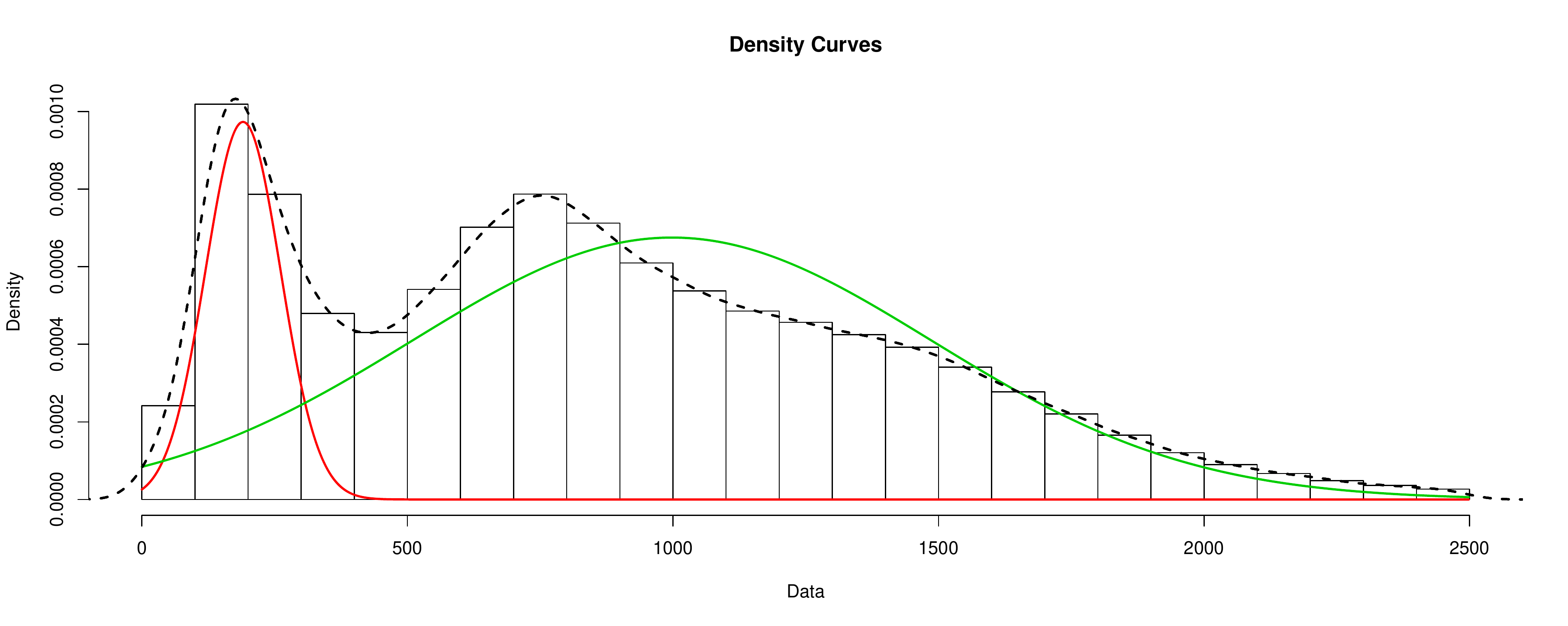} 
\caption{\emph{An example of how the energy consumption density can be represented with the mixture of Gaussian distributions}}
\label{fig:density}
\end{figure}

As an example, Figure \ref{fig:density} demonstrates how consumption variability can be represented as a mixture of densities. As can be seen, we may represent this data with a mixture of Gaussians, although they may differ in size or shape.\footnote{The GMM algorithm is implemented in R in 'mclust' package \citep{scrucca2016mclust}. For the mixture models we utilize a likelihood based estimation procedure.}

\subsection*{Clustering Results}

\begin{figure}[H]
\centering
\includegraphics[width=.45\textwidth]{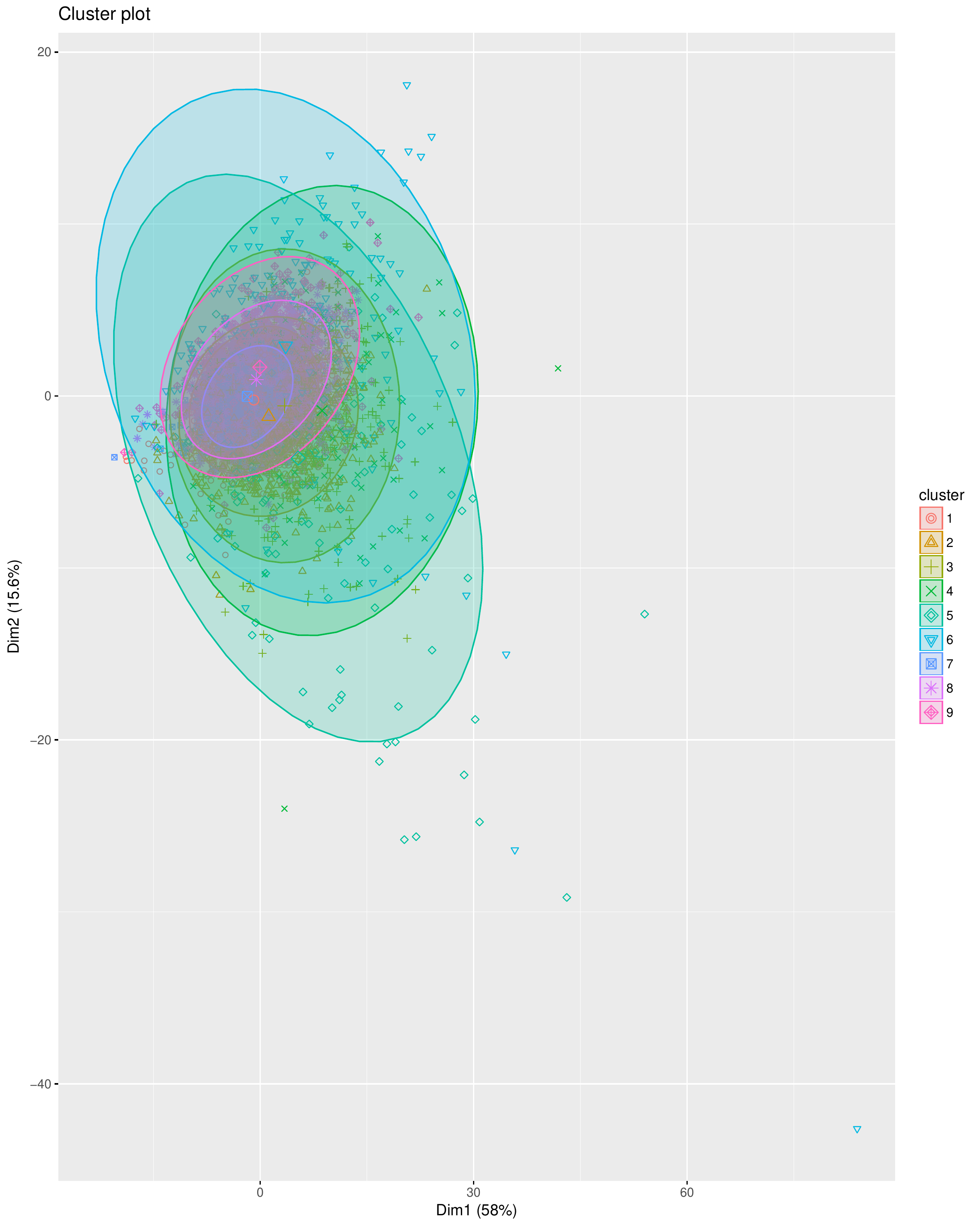}
\includegraphics[width=.45\textwidth]{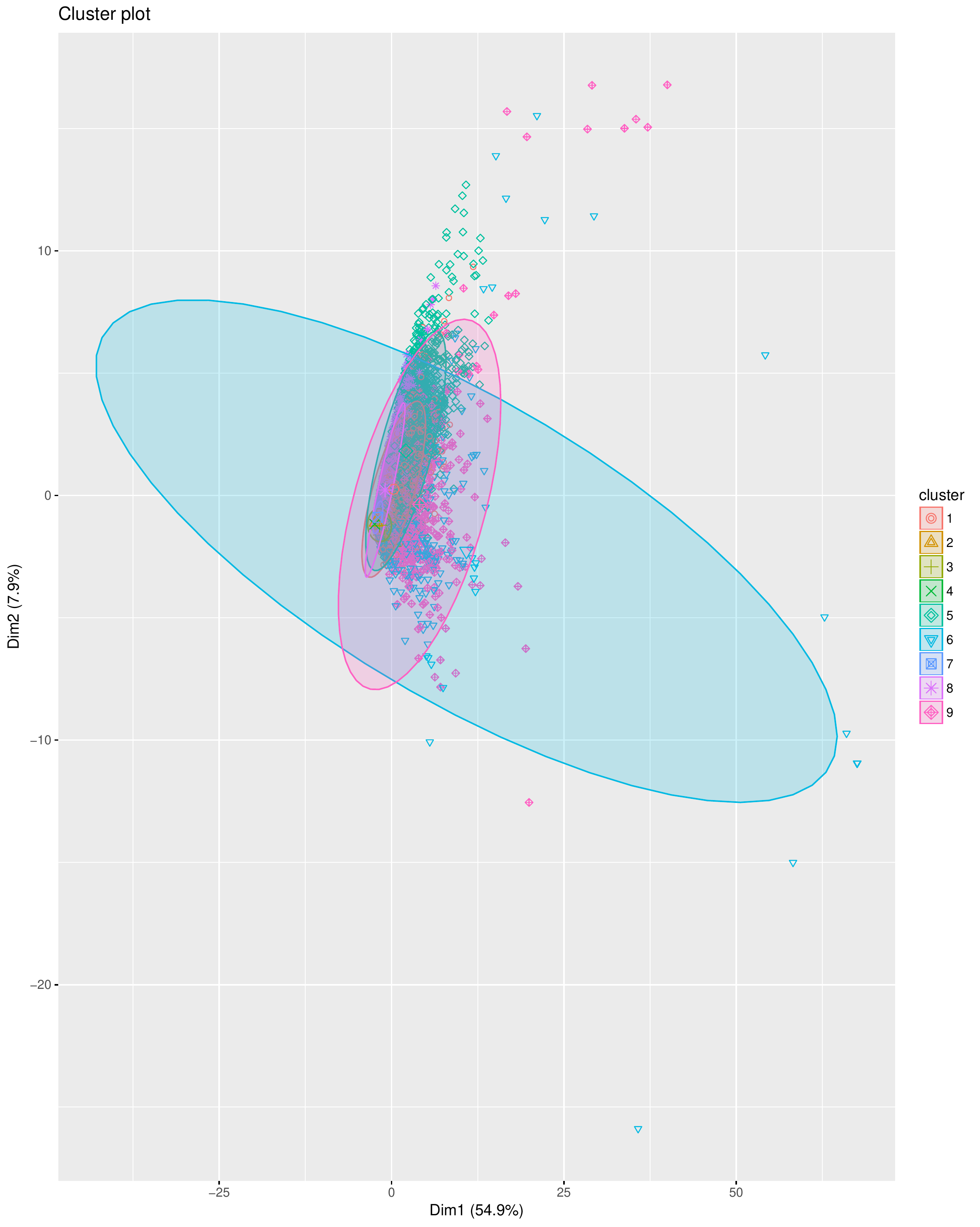}
\caption{\emph{Resulting clusters in high dimensional space}  }
\label{fig:mdg}
\end{figure}

The results of GMM clustering analysis are presented in Figure \ref{fig:mdg} and Table \ref{tab:segment1}. 


As may be observed from Figure \ref{fig:mdg} and Table \ref{tab:segment1}, while we are dealing with different samples we obtained the same number of clustered groups.  However, the key differentiator between the two cluster models is the shape of the Gaussian models used to fit the patterns. While the aggregated sample presents smoother shapes, we see more variation in the disaggregated case (for resulting temporal profiles, see Appendix B). 

In terms of sample allocation to each of the clusters, we are presented with an unbalanced allocation. This is caused by the fact that on average, as we saw in Figure \ref{fig:dec}, energy customers may be alike in their temporal behavior, particularly characterized by morning and evening peaks. In the case of clustering, the less represented groups of patterns are indeed those with lower expected energy consumption, profiles that vary from very low to very high and persistent usage during the day.


\begin{table}[H]
\centering
\begin{center}
\begin{tabular}{ c c c c}
\toprule
Segment\    & \% of total sample (Aggregated patterns)  & \% of total sample (Disaggregated patterns)  \\ [0.5ex] 
\hline \hline
1& 24.0\% & 15.7\% \\ 
2 & 10.6\%  & 14.2\%  \\ 
3 & 5.3\%  & 1.4\% \\ 
4& 0.9\%  & 5.9\% \\ 
5 & 1.9\%   & 20.0\% \\ 
6 & 21.9\% & 3.4\%  \\ 
7 & 15.5\%  & 13.6\% \\ 
8 & 14.4\%   & 22.5\% \\ 
9 & 5.5\%   & 3.4\%\\ 
\bottomrule
\end{tabular}
\end{center}
\caption{ \emph{Results of consumption pattern segmentation using GMM.} }
 \label{tab:segment1}
\end{table}

\section*{Behavior Prediction}

A number of approaches can be used for time series prediction and classification. Initially, it was attempted to forecast the next unit of energy consumption in our data using the standard parametric family of models such as ARIMA, AR, and MA. However, performance was extremely poor and for readability it was decided to omit the details of this analysis here. Instead,  given the greater variability in big data it is proposed  that ability to predict the next half-hour or day of activity may appear troublesome, but as an initial stage the consumer may rather be associated  with a class of known or similar users. In the case presented, we use the labels previously obtained from segmentation of the data. Once again the performance for aggregated and disaggregated samples is then compared. The choice of models was based on their popularity in past research, specifically in the multi-class setting. 

\subsection*{K-Nearest Neighbor}
K-Nearest Neighbor (KNN) is considered one of the simplest classification methods for both binary and multi-class problems. It is particularly useful for problems where the conditional distribution of the outcome variable on the independent variables is unknown \citep{james2013introduction}.  KNN works by taking an input point, $\boldsymbol{x}$, and $K$ points that are in some sense close to it. The points nearby in the feature space can then be used to select an appropriate label. The estimator can be written mathematically as

\begin{equation}
	\widehat{Y}(x)=\frac{1}{K}\sum_{x_i\in N_K(x)}y_i\;,
\end{equation}
where $y_i$ represents the labels of the points in the neighborhood $N_K(x)$ of input point $x$.

\subsection*{Tree-based methods}

The other methods we assess, Random Forest (RF) and Gradient Boosting Trees (GBM), are based solely on decision tree mechanisms. They are differentiated by the approach use to select the best combination of trees and how samples of data are incorporated in the learning process. These methods are especially valuable due to their simplicity in interpretation compared to other machine learning algorithms. They can easily be used for regression and classification type problems and, additionally, to model non-linear relationships.  

The Random Forest algorithm is based on building decision tress on bootstrapped (randomly sub-sampled) data with a smaller subset of randomly sampled predictors at each decision node. A large number of trees is grown until a stopping rule is achieved (e.g. minimum 5 observations in the terminal nodes) and then aggregated for final prediction. An example of the successful use of Random Forest in civil war onset prediction can be found in \cite{muchlinski2015comparing} and \cite{strobl2008conditional}.  

Our implementation of the model is as follows. The input variables are represented by the sequence $\left\{b...B\right\}$ which is a combination of half-hourly readings. The model draws bootstrap samples, Z, from the training set, and random forest trees are built using a combination of predictors that are responsible for the split of these trees. Once a number of tree classifiers have been generated, we take the average among all and form a single classifier.  Output is represented by  $\left\{T_{b}\right\}_1^B$ The class is then predicted for the unseen data (test set) through the majority vote that selects the best performing trees :

 \begin{equation}
 \widehat{C_rf^B(x)} = majority\ vote \left\{\widehat{C_b(x)}\right\}_1^B
 \end{equation}

An alternative tree algorithm known as Gradient Boosting, first used to tackle classification problems, is now widely used for regression as well \cite{friedman2001elements}. Like Random Forest, the gradient boosting algorithm takes advantage of both weak and strong classifiers. By weak, we mean classifiers that bring a prediction which is slightly better or just the same as a random guess. Unlike Random Forest where at each iteration we are training a different solution, in the Gradient Boosting model we are updating the solution of the already trained model as more samples are taken. The trees are, therefore, updated at each iteration to obtain more powerful classifiers. 

In boosting models, we first assign the weights  $w_i = \frac{1}{N} $ to each of our training observations that include both input and output variables, with $N$ being the total number of observations. We then iterate the process  $F$ times during which we fit the classifier $G_f(x)$ using the observation weights. The observations which were misclassified at the previous stage are assigned greater weights, so at each iteration we give more importance to those observations that were harder to classify initially. We calculate the error associated with each model fit as

 \begin{equation}
 	 e_f= \frac{\sum _{i \epsilon N_i} w_i I (y_i \neq G_f(x_i))} {\sum _{i \epsilon N_i} w_i}
 \end{equation}

Those with the highest error are assigned an increase to their weights using the factor of $\exp {\gamma_f}$. The final output $G(x)$ is based on continuous iterations of model fit using re-weighted observations until the error rate is minimized.

\section*{Results}

The tables below report overall accuracy and kappa values for each of the models were used to predict the data segment. Entries for `Accuracy' report the overall prediction power of the model including both true positives and true negatives over total of true and false positives and negatives. The Kappa statistic is used for the evaluation of classifiers by comparing the observed accuracy of prediction with that of a random chance. The optimal parameters were obtained using ten-fold cross-validation. The results are followed  by confusion tables that represent the ratio of observed versus predicted class.

\subsection*{Aggregated Results}

\begin{table}[H]
\centering
\begin{center}
\begin{tabular}{ c c c c c}
\toprule
 Model\     & Accuracy & Kappa \\ [0.5ex] 
\hline \hline
K-Nearest Neighbor & 23\%  & 0.14   \\ 
Gradient Boosting Trees & 37\% & 0.29  \\ 
Random Forest & 40\%&  0.29  \\ 
\bottomrule
\end{tabular}
\end{center}
\caption{ \emph{Results of multi-class prediction on aggregated sample}.}
 \label{tab:prediction}
\end{table}

\begin{figure}[H]
\centering
\includegraphics[width=0.9\textwidth]{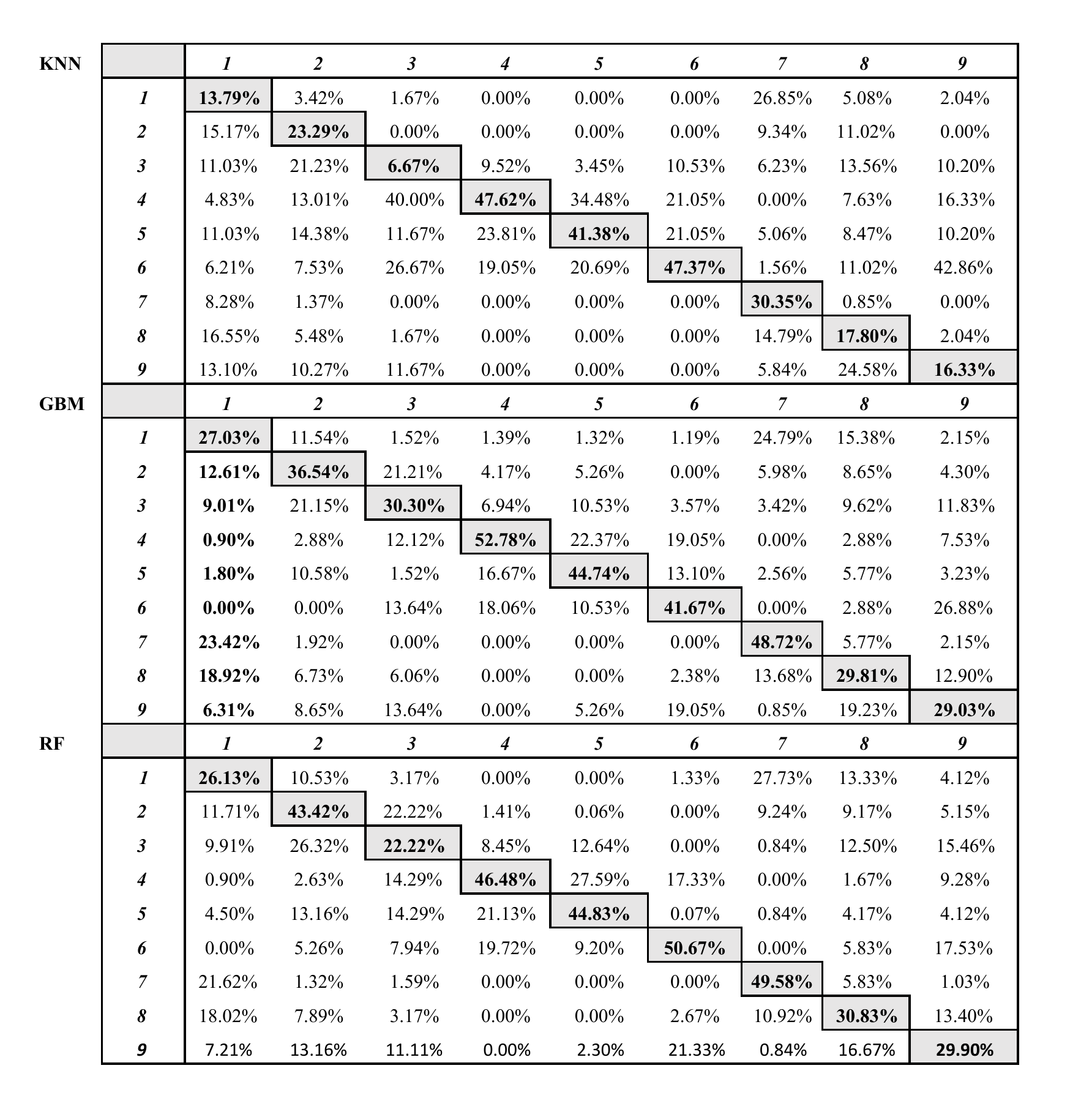}
\caption{\emph{Confusion matrix reporting the correspondence between predicted (rows) vs observed class (columns)}.}
\label{fig:badconfusion}
\end{figure}

\subsection*{Results on disaggregated sample}

\begin{table}[H]
\centering
\begin{center}
\begin{tabular}{ c c c c c}
\toprule
 Model\     & Accuracy & Kappa   \\ [0.5ex] 
\hline \hline
K-Nearest Neighbor & 65\% & 0.58   \\ 
Gradient Boosting Trees & 80\% &  0.73  \\ 
Random Forest & 79\% & 0.75 \\ 
\bottomrule
\end{tabular}
\end{center}
\caption{ \emph{Results of multi class prediction on disaggregated sample} }
 \label{tab:prediction_disag}
\end{table}

\begin{figure}[H]
\centering
\includegraphics[width=0.9\textwidth]{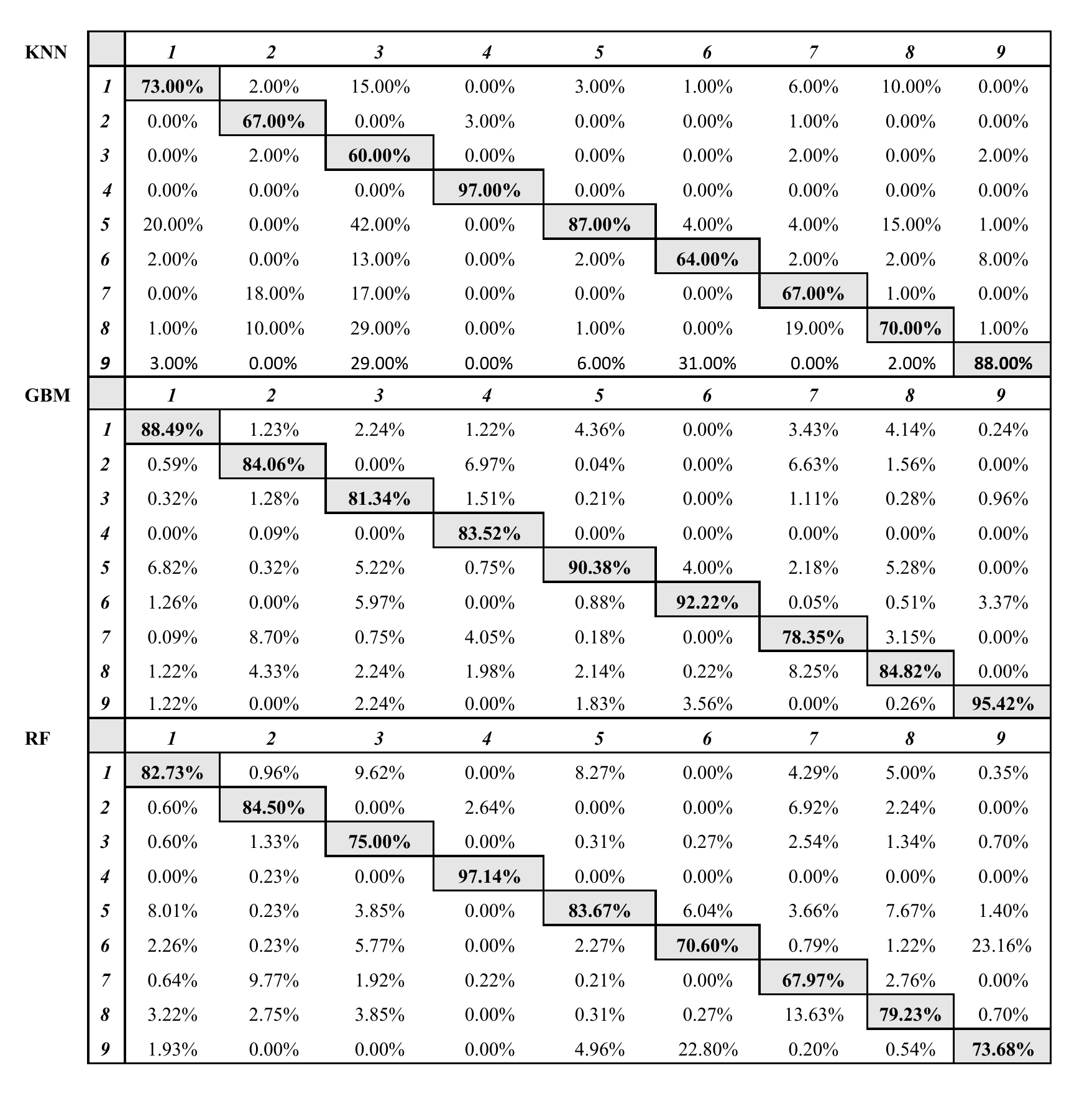}
\caption{\emph{Confusion matrix reporting the correspondence between predicted (rows) vs observed class (columns).}  }
\label{fig:confusion}
\end{figure}

As observed from the confusion tables (Figures \ref{fig:badconfusion} and \ref{fig:confusion}), the prediction methods show differential performance across clusters. One of the immediate observations is the difference in performance when considering aggregated versus disaggregated analysis (Tables \ref{tab:prediction} and \ref{tab:prediction_disag}). Aggregated models are associated with higher misclassification rates, suggesting that by aggregating we have lost essential dynamics that contribute to identifiable patterns. 

While RF and GBM tend to perform better on average, KNN showed higher accuracy in some classes. This is possibly related to different `bias-variance' trade off for each of the tree models. While boosting aims to reduce the bias by taking the average of predictive performance among the estimated models, Random Forest fundamentally searches for a solution that reduces variance by imposing a strict structure of reducing the number of predictors at each split of the tree. 

Often, the classes that are better represented in the data may be associated with better performance as there is more data available for the training. In our case, this had no implication on performance. Classes with smaller number of observations were more easily differentiated, while the bigger ones showed higher levels of misclassification.

\section*{Discussion and Conclusions}

In this paper the analysis that can be performed on time series associated with substantial levels of variability across a large number of data-streams was presented. It was demonstrated that such data can be  meaningfully clustered using Gaussian Mixture Models. The paper suggests a possible strategy for prediction and characterization of temporal profiles. One of the arising challenges is the effect of aggregation on prediction performance.

It was shown that both segmentation and predictive algorithms tend to work differently depending on whether we looked at aggregate or disaggregate samples. For prediction in particular, we show that using aggregated data records leads to much higher rates of misclassification, while the most granular data can be classified and predicted with more certainty. 

Compared to Random Forest, in practice some classifications may be better performed using Gradient Boosting trees \citep{friedman2001elements}. However, the performance may be at the cost of over-fitting the data. Nevertheless, what is observed is rather a mixture of performances with each method winning or losing for different prediction class. This may be related to the essential `bias-variance' trade-off that is worked differently by each model. While boosting aims to reduce the bias by taking the average of predictive performance among the estimated models, Random Forest fundamentally searches for the solution that reduces the variance by imposing a strict structure of reducing the number of predictors at each split of the tree. 

In previous research, model stacking has shown an improvement in performance \citep{rokach2010ensemble, dietterich2000ensemble}. In future work we will evaluate a mix of the models as an ensemble to improve overall predictive performance.

\clearpage \singlespacing
\bibliographystyle{apsr}
\bibliography{06-08}{}

\newpage
\section*{Appendix}

\begin{figure}[H]
\centering
\includegraphics[width=0.8\textwidth]{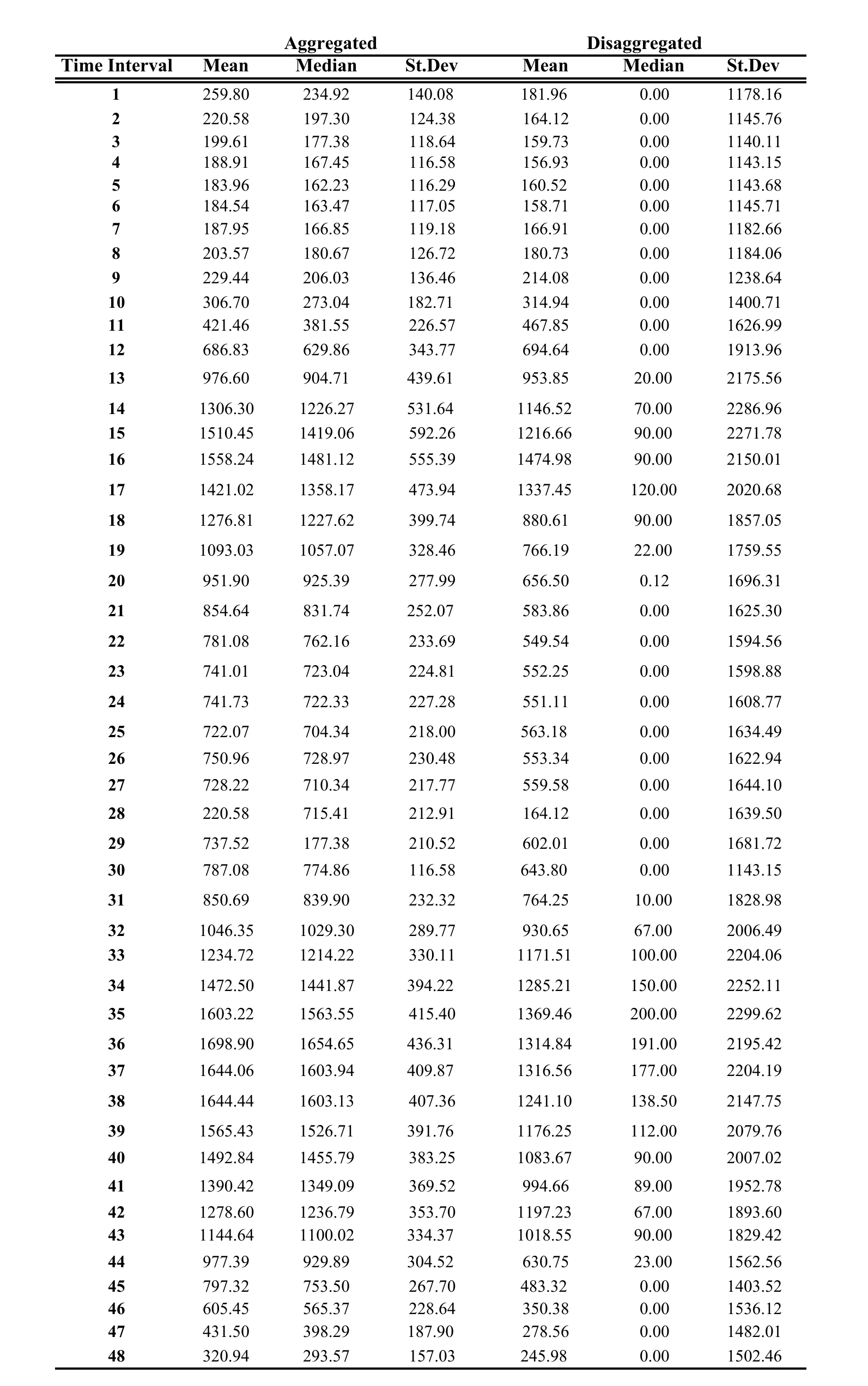}
\caption{\emph{Descriptive statistics for the samples we used for the experiments}  }
\label{fig:desc}
\end{figure}

\appendix{Appendix (A)}
Figure \ref{fig:desc} presents the description of the data used for the analysis: aggregated and disaggregated sample.

\newpage
\appendix{Appendix (B)}

\begin{figure}[H]
\centering
\includegraphics[width=0.6\textwidth, angle =90]{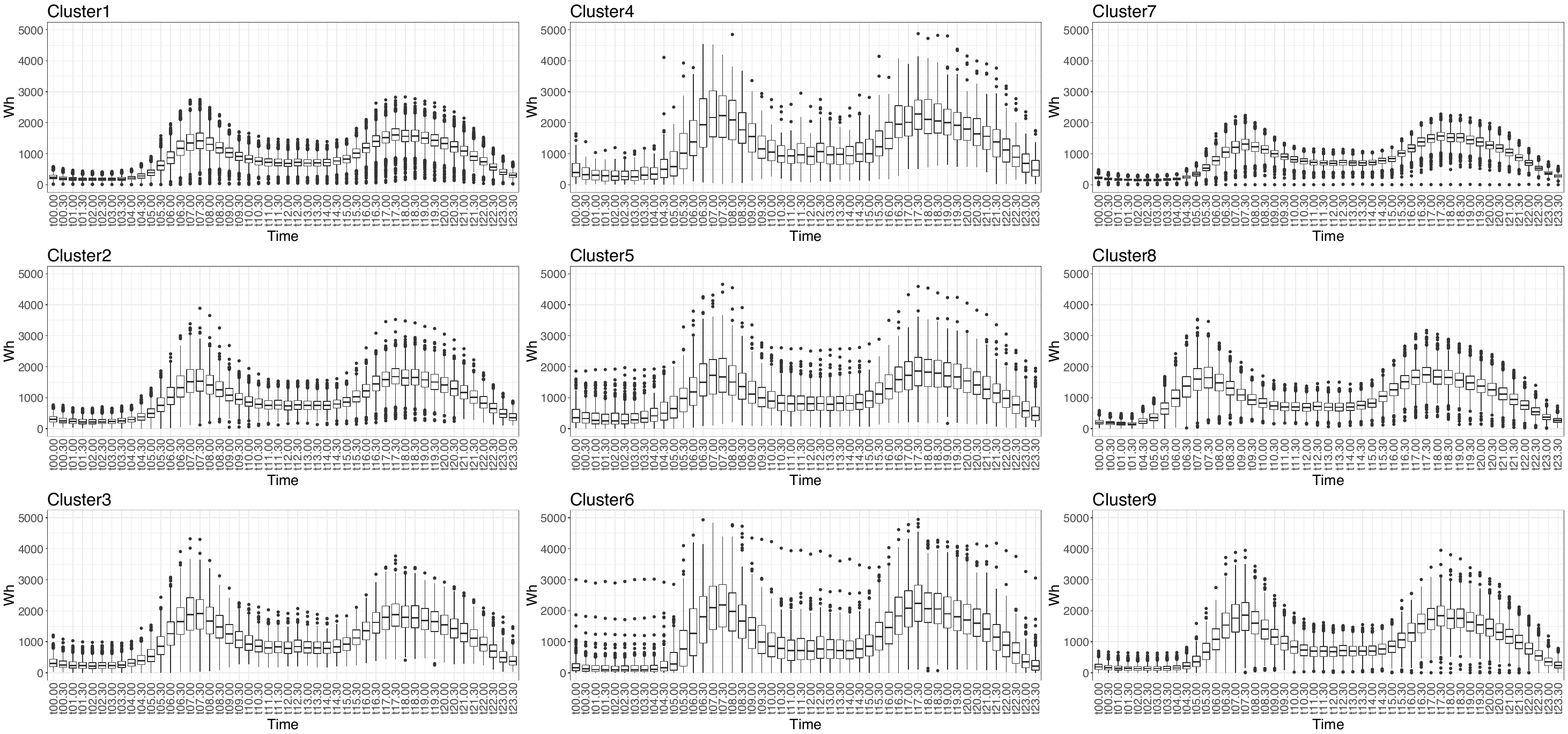}
\caption{\emph{Clusters observed on aggregated sample}  }

\label{fig:agcl}
\end{figure}

\begin{figure}[H]
\centering
\includegraphics[width=0.75\textwidth]{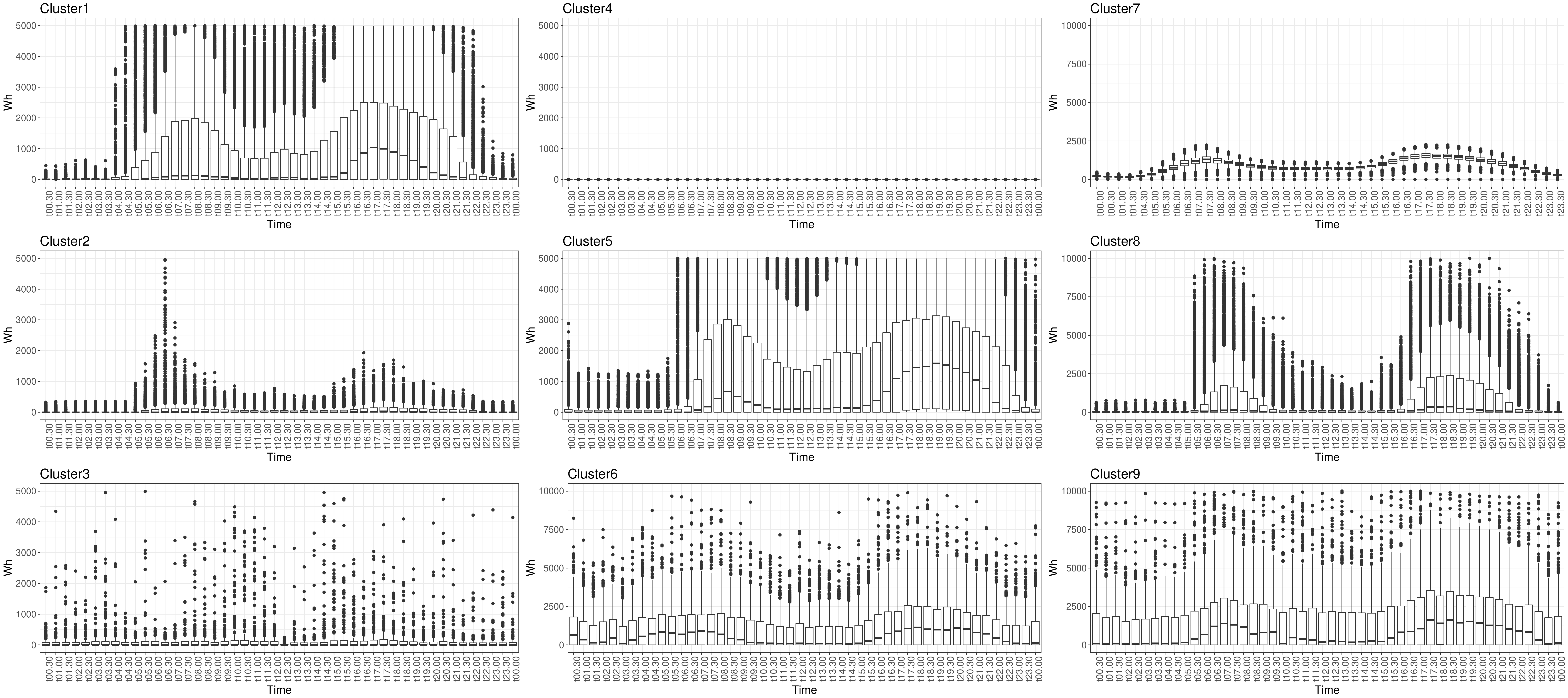}
\caption{\emph{Clusters observed on disaggregated sample}  }

\label{fig:discl}
\end{figure}

Figures \ref{fig:agcl}  and \ref{fig:discl} present the shapes and variation in the resulting clusters using the GMM model. As can be seen the number of clusters was defined as identical, however, the shape of aggregated clusters is far more smoother then those of the disaggregated sample. This fundamental difference may have had a direct implication for the predictability of aggregated clusters as the differentiation on the aggregated level may be more challenging as essential dynamics that distinguish the patterns were collapsed during the averaging of energy consumption.

\end {document}